\documentclass[aps,pra,reprint,superscriptaddress]{revtex4-2}

\usepackage{xcolor}
\usepackage{graphicx}
\usepackage{amsmath,ulem}

\begin{document}

\title{Enhanced spatiotemporal optical vortices and vortex chains from Hermite-Gauss modes with a tilted pulse front}
\author{Miguel A. Porras}%
\email{miguelangel.porras@upm.es}
\affiliation{Grupo de Sistemas Complejos, ETSIME, Universidad Politécnica de Madrid, Rios Rosas 21, 28003 Madrid, Spain}
\author{Spencer W. Jolly}
\email{spencer.jolly@ulb.be}
\affiliation{Service OPERA-Photonique, Université libre de Bruxelles (ULB), Brussels, Belgium}

\date{\today}

\begin{abstract}
Hermite-Gaussian (HG) beams are standard modes delivered by continuous or pulsed lasers systems, and pulse-front tilt is one of the most common, detrimental or beneficial, spatiotemporal couplings affecting ultrashort pulses. Combining them, we show that focusing a pulsed HG beam with a tilt generates an elliptical spatiotemporal optical vortex (STOV), or a chain of them. The elliptical STOVs differ from standard STOVs in an additional spatial chirp that is manifested as a wave front temporal rotation, and results in an enhanced transverse orbital angular momentum. The longitudinal field is significantly larger than that of normal STOVs, and may also take the form of an elliptical STOV. Our concept greatly simplifies previous arrangements for the generation of STOVs, whose additional features make them attractive for improving their applications in electron trapping and acceleration, or as driving fields for the generation of higher-order harmonics and other interactions with matter.
\end{abstract}

\maketitle

\section{Introduction}

Spatiotemporal optical vortices (STOVs), also called spatiotemporal orbital angular momentum (ST-OAM) beams and transverse OAM beams, are fascinating optical fields with a rich spatiotemporal and topological structure, which are now at the core of the research in structured light, and particularly light carrying OAM. The distinctive property of STOVs is a phase singularity on a line perpendicular to the direction of propagation, which creates a dislocation in the electric field oscillations. If the singularity is surrounded by an elliptical ring of light, such a STOV field carries a well-defined amount of transverse OAM of rotation about itself. 

STOVs were first described theoretically~\cite{sukhorukov05,bliokh12}, and later observed in a nonlinear filamentation experiment~\cite{jhajj16}. The explosion of interest in STOVs originated after their controlled generation~\cite{hancock19,chong20,chen22} in free space or air using standard optics such as adapted $4f$ pulse shapers to include spiral or tilted $\pi$-step phase plates~\cite{hancock19}, or spatial light modulators~\cite{chong20,huang24}. Alternative proposals include the use of photonic crystal slabs~\cite{WANG_OPTICA_2021} or metasurfaces~\cite{HUANG_PRB_2023}, and more recently STOVs have been realized in experiments with meta-optics~\cite{huo24}. The experimental generation has been accompanied by specific characterization techniques~\cite{chen21,gui22}, actual or foreseen applications as information carriers~\cite{huang24}, for electron trapping and acceleration~\cite{sunF24}, and by a great deal of theoretical work to understand their intriguing propagation properties~\cite{huang21,huang22,chenj22,rui22,porras23-1,hyde23,porras25} and the nature of their transverse OAM~\cite{porras23-4,bliokh23,porras24-1,porras24-2}. In the meantime, STOVs have been employed to up-convert their structure into second harmonic, sum frequency, third harmonic, and high-order harmonic fields in the extreme ultraviolet~\cite{guanG21,gao23,wang23,fang21,dong24,rodrigo24}. 	

\begin{figure}
	\centering
	\includegraphics[width=86mm]{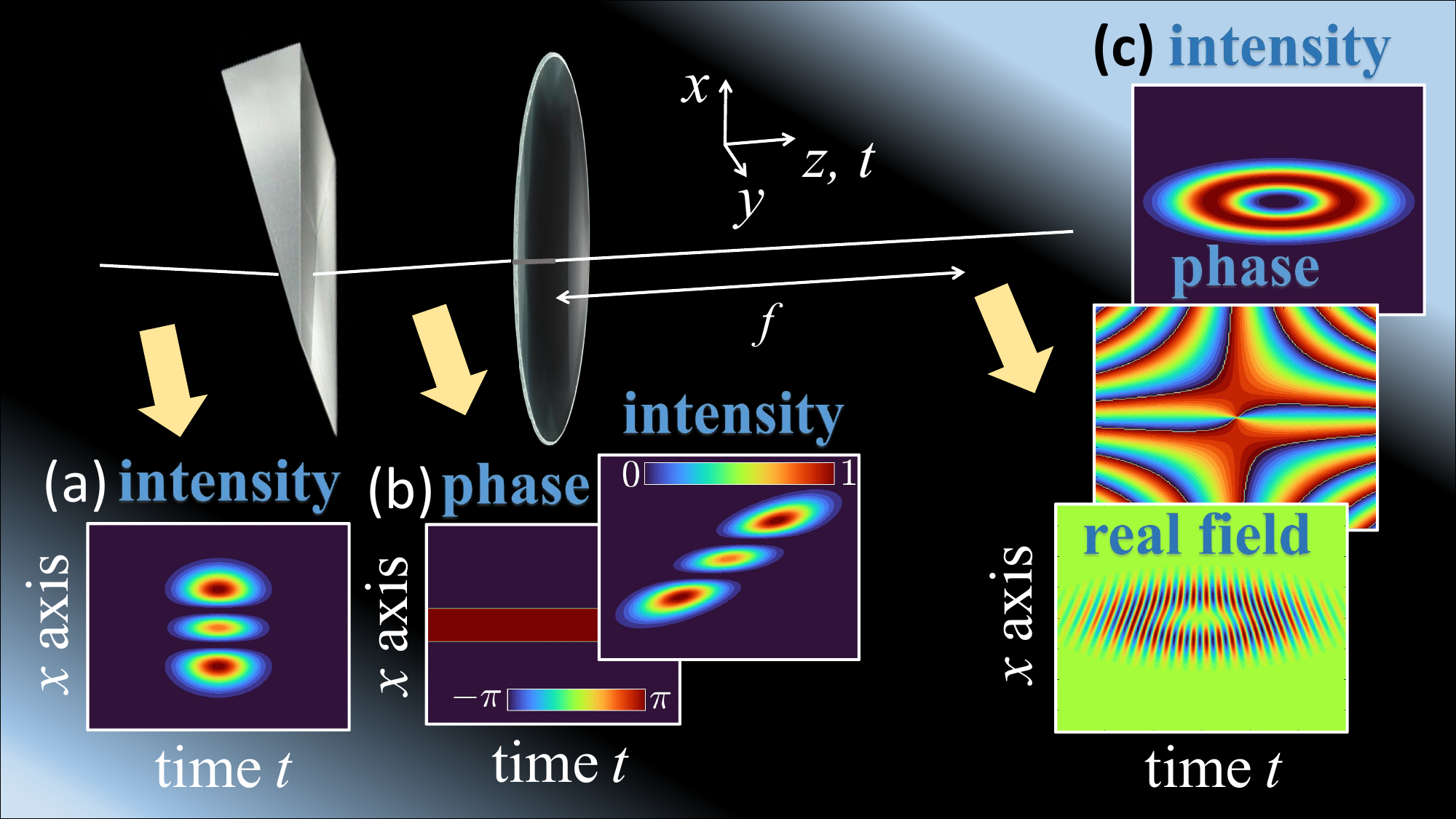}
	\caption{Illustration of the optics that generates a E-STOV, or a chain of STOVs. (a) Input pulsed HG mode. (b) tilted-pulse HG mode. (c) Focal STOV with added spatial chirp produced with appropriate input positive or negative tilt. The topological charge is $l=\pm n$, where $n$ is the Hermite polynomial order.}
	\label{fig:1}
\end{figure}

In this paper we introduce a new class of STOVs featuring a significant enhancement of all relevant properties of standard STOVs that have made them attractive and useful. In addition, generating these new STOVs seems simpler than generating standard STOVs. In Fig.~\ref{fig:1}, a Hermite-Gauss (HG) mode from a pulsed laser source (a) is tilted (b) and focused (c). With adequate, positive or negative tilt, an elliptical STOV of topological charge $l=\pm n$ of absolute value equal to the order $n$ of the Hermite polynomial is formed at the focal plane. Whereas a standard STOV presents a weak spatial chirp, these STOVs present a significant, added spatial chirp. The tilted HG pulse on the lens here differs substantially from the spatiotemporal HG pulses that focus to standard STOVs \cite{porras24-1,porras25}. As in Fig. \ref{fig:1}, the HG pulse is tilted without tilting each particular lobe, which requires a simple wedge, while in spatiotemporal HG pulses \cite{porras24-1,porras25} the tilt as a whole is accompanied by an opposite tilt of each lobe. The generation of such a structure requires a $4f$ pulse shaper with a spiral phase plate at the common focus \cite{hancock19}. Our technique can then be said to be simpler, as it simply combines the ever-present HG modes and tilting, the simplest spatiotemporal structuring of light.

The transverse angular momentum (OAM) per unit energy in our case is more than twice the OAM of normal STOVs, and the longitudinal component is about twice more pronounced. Interestingly, even in the absence of a STOV in the transversal field component ($n=0$), a STOV is still present in the longitudinal field component. Such a STOV in a longitudinal field has not previously been described. To distinguish these STOVs with enhanced properties from standard STOVs, we will call them enhanced STOVs (E-STOVs). 

Switching the tilt to lower or higher absolute values than the above optimal value, which is possible by simply rotating the prism in Fig.~\ref{fig:1}, the multiply charged STOV splits into an array of $n$ spatially or temporally sorted vortices of charges $\pm 1$, with also added spatial chirp and respective lower and higher values of the OAM. These vortex chains may be of interest in information-related applications such as information storage and transmission~\cite{huang24}. Note importantly, that the separation of vortices along different dimensions is an emergent property of the same simple tilted HG pulse that is subsequently focused, and not a construction at the focal plane, as in~\cite{Gu:23}.

The theory is developed here in the paraxial approximation and for quasimonochromatic (many-cycle) pulses. This approximation is accurate in majority of the scenarios where STOVs are created and used, which involve durations of several tens of femtoseconds and focusing with small numerical apertures. Within these approximations, closed-form analytical expressions are provided for the E-STOVs and vortex chains while they are formed upon focusing and disintegrate, as normal STOVs do.  

In Sec. \ref{sec:main} we derive our main result. The spatiotemporal and spatiospectral structure, and the enhanced spatial shirp of E-STOVs and STOV chains are analyzed in Sec. \ref{sec:analysis}. In Sec. \ref{sec:OAM} we evaluate the enhanced transverse OAM of E-STOVs as due to pulse front and wave front rotation contributions. Finally, the associated, enhanced longitudinal field is evaluated in Sec. \ref{sec:axial}.

\section{Analytical field for focused Hermite-Gaussian mode with tilted pulse-front}\label{sec:main}

As is well-known,\cite{akturk04} the complex envelope of a generic tilted pulse (TP) at a plane, e.g., $z=0$, perpendicular to the propagation direction is of the form $\psi(x,t)=\psi_x(x)\psi_t(t-px)$, where the parameter $p$ determines the tilt angle in space $(x,z)$ with respect to the $x$ direction as $\theta=\tan^{-1}(cp)$ if the pulse travels in free space at velocity $c$. An arbitrary factorized profile along the $y$ direction is assumed, e.g., a Gaussian profile, that remains factorized upon propagation and hence will not be written. We take the TP HG mode specified by 
\begin{equation}\label{eq:tilted}
    \psi_x(x)= \frac{1}{2^n}H_n\left(\frac{\sqrt{2}x}{X_0}\right) e^{-\frac{x^2}{X_0^2}},\quad \psi_t(t)=e^{-\frac{t^2}{t_0^2}},
\end{equation}
where $X_0$ and $t_0$ specify the width and duration, and $H_n$ is the Hermite polynomial of order $n$. The non-tilted HG mode is illustrated in Fig. \ref{fig:1}(a). The TP HG mode  $\psi(x,t)=\psi_x(x)\psi_t(t-px)$ can be produced upon passage through a single diffraction grating or a wedge, as drawn in Fig. \ref{fig:1}(b) for its intensity and phase. If the TP HG mode  illuminates an ideal focusing element of focal length $f$ at $z=0$, the focused field can be evaluated in terms of the Fresnel diffraction integral as
\begin{align}\label{eq:Fresnel}
    \begin{split}
    &\psi(x,t',z) =\\
    &\sqrt{\frac{k_0}{2i\pi z}} \int_{-\infty}^{\infty} dx' \psi_x(x')\psi_t(t'-px') e^{\frac{-ik_0x'^2}{2f}}e^{\frac{ik_0}{2z}(x-x')^2},
    \end{split}
\end{align}
where $k_0=\omega_0/c$ is the propagation constant, $\omega_0$ is the carrier frequency, and $t'=t-z/c$ is the local time. The Fresnel integral involves the paraxial approximation, which requires ${\rm NA}= \sin [\tan^{-1} (X_0/f)]\simeq X_0/f\ll 1$. Its use for quasimonochromatic (many-cycle) pulses is justified in Appendix~\ref{sec:appA} starting from the wave equation for this type of light. Also, Eq. (\ref{eq:Fresnel}) neglects the pulse front curvature imparted by the focusing system. As detailed in Appendix~\ref{sec:appA}, pulse front curvature effects are negligible if the input illumination satisfies $X_0^2/2cf\ll t_0$. In practice, one may wish to produce a focal E-STOV or STOV chain of certain scale $x_0=2f/(k_0X_0)$ (focal Gaussian size). Then, the condition $X_0^2/2cf\ll t_0$ of small or negligible pulse front curvature effects reads as the condition $f\ll ct_0k_0^2x_0^2/2$ for the focal length, and then the choice $X_0=2f/k_0x_0$ for the input size. All examples in this paper satisfy the condition of negligible pulse front curvature effects. An example of distortions when $f$ approaches $ct_0k_0^2x_0^2/2$ is given in Appendix \ref{sec:appA}. 

We note that Ref.~\onlinecite{porras24-1} considers the particular case with $n=0$ that focuses to a rotating pulse without any spatiotemporal vortex. The same Ref.~\onlinecite{porras24-1} also considers a lens illumination expressed by a spatiotemporal Hermite polynomial, but it does not have the form of a tilted pulse. Such an illumination is the output of a $4f$ pulse shaper with a spiral phase plate at the Fourier plane \cite{hancock19}, and focuses to a standard STOV \cite{hancock19,porras24-1}. Instead, in this case, we only need a standard Hermite-Gaussian beam and a single diffraction grating or a wedge. 

As detailed in Appendix~\ref{sec:appA}, the focused field at any distance $z$ from the lens provided by Fresnel diffraction integral (\ref{eq:Fresnel}) can be expressed in closed analytical form as
\begin{align}\label{eq:main}
\begin{split}
    &\psi(x,t',z)=\sqrt{\frac{k_0}{2iz}}e^{\frac{ik_0x^2}{2z}}X_{0,\rm eff}\, e^{X_{0,\rm eff}^2\left(\frac{pt'}{t_0^2}-\frac{ik_0x}{2z}\right)^2} e^{-\frac{t'^2}{t_0^2}} \\
    &\times\left(1-\frac{2X_{0,\rm eff}^2}{X_0^2}\right)^{\frac{n}{2}} \frac{1}{2^n} H_n\left[\frac{\sqrt{2}}{X_0}\frac{X_{0,\rm eff}^2\left(\frac{pt'}{t_0^2}-\frac{ik_0x}{2z}\right)}{\sqrt{1-\frac{2X_{0,\rm eff}^2}{X_0^2}}}\right],
\end{split}
\end{align}
where
\begin{equation}\label{eq:xeff}
    \frac{1}{X_{0,\rm eff}^2} = \frac{1}{X_0^2} + \frac{p^2}{t_0^2} - \frac{ik_0}{2}\left(\frac{1}{z}-\frac{1}{f}\right).
\end{equation}

\section{Enhanced STOVs and STOV chains}\label{sec:analysis}

The plane of greatest interest is the focal plane. At $z=f$, $X_{0,\rm eff}$ becomes real. It is then convenient to replace it with the more physical
\begin{equation}\label{eq:parameters}
    x_{0,\rm eff} = x_0 \sqrt{1+\frac{p^2X_0^2}{t_0^2}},\quad t_{0,\rm eff} =t_0\sqrt{1+\frac{p^2X_0^2}{t_0^2}}
\end{equation}
effective width and duration, and the focal Gaussian width $x_0=2f/(k_0X_0)$ in absence of tilt. Then Eq. (\ref{eq:main}) becomes
\begin{align}\label{eq:focal}
\begin{split}
    &\psi(x,t',f)=\sqrt{\frac{-iX_0t_0}{x_{0,\rm eff}t_{0,\rm eff}}} e^{\frac{ik_0x^2}{2f}}e^{-\frac{x^2}{x_{0,\rm eff}^2}}\left(\frac{\frac{p^2X_0^2}{t_0^2}-1}{\frac{p^2X_0^2}{t_0^2}+1}\right)^{\frac{n}{2}} \\
    &\times\frac{1}{2^n}H_n\left[\sqrt{\frac{2}{\frac{p^2X_0^2}{t_0^2}-1}}\left(\frac{pX_0}{t_0} \frac{t'}{t_{0,\rm eff}}-i\frac{x}{x_{0,\rm eff}}\right)\right] \\
    &\times e^{-\frac{t'^2}{t_{0,\rm eff}^2}} e^{-2i\frac{pX_0}{t_0} \frac{xt'}{x_{0,\rm eff}t_{0,\rm eff}}},
\end{split}
\end{align}

There are two "magic" values of the tilt parameter, $p= \pm t_0/X_0$, with which the last factor in the first row cancels out all terms of the Hermite polynomial except the highest power term, and Eq. (\ref{eq:focal}) further simplifies to the elliptical STOV
\begin{align}\label{eq:elliptical}
\begin{split}
    \psi(x,t',f)&=\sqrt{\frac{-iX_0}{2x_0}}e^{\frac{ik_0x^2}{2f}}e^{-\frac{x^2}{2x_0^2}}(\pm 1)^n\left(\frac{t'}{\sqrt{2}t_0}\mp i\frac{x}{\sqrt{2}x_0}\right)^n \\
    &\times e^{-\frac{t'^2}{2 t_0^2}}  e^{\mp i\frac{xt'}{x_0t_0}},
\end{split}
\end{align}
of topological charge $l=\pm n$, with an added spatial chirp represented by the factor $e^{\mp i xt'/x_0t_0}$, or E-STOV. The ellipticity is the result of the equality of the widths and durations, $\sqrt{2}x_0$ and $\sqrt{2}t_0$, of the STOV core and the Gaussian envelopes. The shape of the intensity, phase and real field are illustrated in Fig. \ref{fig:1}(c). Compared to a normal STOV, the added spatial chirp is observed in the bottom panel of \ref{fig:1}(c) as significant opposite tilts of the wave fronts at negative and positive times, which amounts to a wave front rotation in time~\cite{quere14,auguste16}, which is absent in normal STOVs. The residual wave front curvature factor $e^{ik_0x^2/2f}$ observable in the middle panel of Fig. \ref{fig:1}(b) is common to standard STOVs formed upon focusing \cite{porras24-1}, and is increasingly negligible as focusing is stronger.

\begin{figure}
	\centering
	\includegraphics[width=86mm]{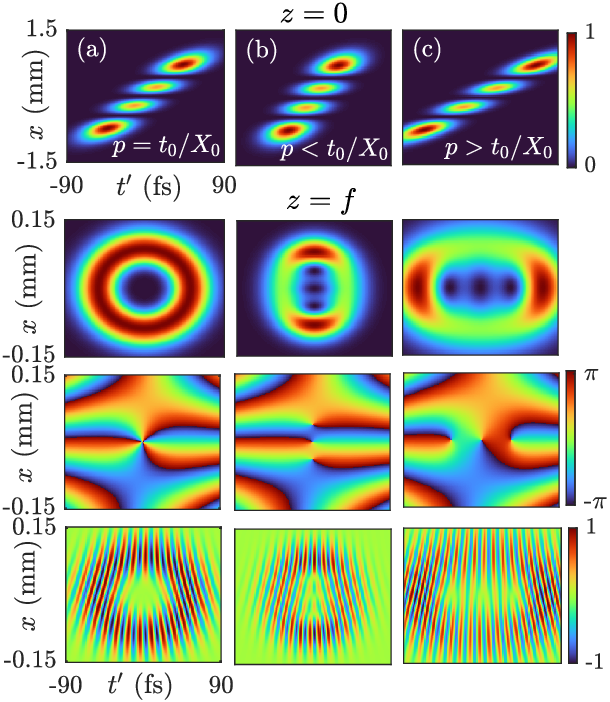}
	\caption{First row: Intensity of TP HG modes in front of the lens. The second, third and fourth rows represent the intensity, phase and real field (the carrier frequency is reduced by a factor of 5 for easier visibility) at the focal plane. The columns (a), (b) and (c) correspond to respective tilt parameters $p=t_0/X_0$ producing an E-STOV,  $(2/3)t_0/X_0$, and $p=(3/2) t_0/X_0$ producing STOV chains. Data: $\lambda_0= 800$ nm ($\omega_0=2.366$ rad/fs), $X_0=0.5$ mm, $t_0=30$ fs, $f=100$ mm, and $n=3$, resulting in with topological charge $l=+n=3$.}
	\label{fig:2}
\end{figure}

All these properties are seen in more detail in Fig. \ref{fig:2} for realistic values of the parameters and a higher topological charge, together with the focal field with values of $p$ other than $p=\pm t_0/X_0$, as given by the more general Eq. (\ref{eq:focal}). For $|p|< t_0/X_0$, the square root inside the Hermite polynomial is pure imaginary; hence the term with $t'$ becomes imaginary and the term with $x$ is real. Since the $n$ zeroes of the Hermite polynomials, determining the location of the $n$ vortices, are all real, these are located at $t'=0$ as $n$ vortices of unit charge, i.e., sorted along the $x$ axis, as in Fig. \ref{fig:2}(b). The opposite occurs with $|p|> t_0/X_0$, where the $n$ vortices of unit charge are sorted temporally, as in Fig. \ref{fig:2}(c). It also follows from Eqs. (\ref{eq:focal}) and (\ref{eq:parameters}), that the spatial chirp $-2p(X_0/t_0)/[x_0t_0(1+p^2X_0^2/t_0^2)]$ is maximum in magnitude for $|p|=t_0/X_0$, becoming negligible for $p$ approaching zero or very large values. Indeed, the intensity patterns approach in these two limits a $x$-sorted multi-hump pulse or a $t'$-sorted multihump pulse without spatial chirp, with the vortices degenerated into $n$ horizontal or vertical zero-intensity lines.

A deeper understanding of these E-STOVs and STOV chains is provided by their spatially resolved spectrum, or spatiospectrum for short. The temporal Fourier transform $\hat \psi= (1/2\pi) \int_{-\infty}^{\infty} \psi e^{i\Omega t'} dt'$, where $\Omega=\omega-\omega_0$, of Eq. (\ref{eq:focal}) can be evaluated from integral 7.374.8 of Ref.~\onlinecite{gradshteyn07}. Some long algebra and changes of variables leads to the particularly simple expression
\begin{align}\label{eq:spatiospectrum}
\begin{split}
\hat{\psi}(x,\Omega,f)=&\sqrt{\frac{-iX_0}{x_0}}e^{\frac{ik_0x^2}{2f}}e^{\frac{-\Omega^2}{\Omega_0^2}} e^{-\frac{(x-b\Omega)^2}{x_0^2}} \\
&\times\frac{1}{2^n} H_n\left[\frac{\sqrt{2}(x-b\Omega)}{x_0}\right],
\end{split}
\end{align}
where $b= (f/k_0)p$, $\Omega_0=2/t_0$ is the Gaussian bandwidth, and a irrelevant $i^n/(\sqrt{\pi}\Omega_0)$ factor is omitted. It is not surprising that the spatiospectrum resembles the input TP HG mode given the symmetric roles of $x$ and $t'$ in Eq. (\ref{eq:focal}), with the difference that the spectrum as a whole is not tilted but the lobes are tilted. Equation (\ref{eq:spatiospectrum}) is immediately identified with a standard HG beam at each frequency $\omega_0+\Omega$ to which the spatial chirp has been directly introduced in the frequency domain with the replacement $x\rightarrow x-b\Omega$. This is consistent with the well-known connection between angular dispersion (pulse-front tilt) and spatial chirp when focusing ultrashort pulses. In fact, the propagated field at any $z$ in Eq. (\ref{eq:main}) could be re-obtained from well-known propagation formulas of the HG mode in (\ref{eq:spatiospectrum}) at the focal plane and an inverse temporal-frequency Fourier transform. With the frequency-domain formulation, properties relevant to tighter focus and shorter pulses can be retained, such as the pulse front curvature, as detailed in Appendix~\ref{sec:appB}.

\begin{figure}
\centering
 \includegraphics[width=86mm]{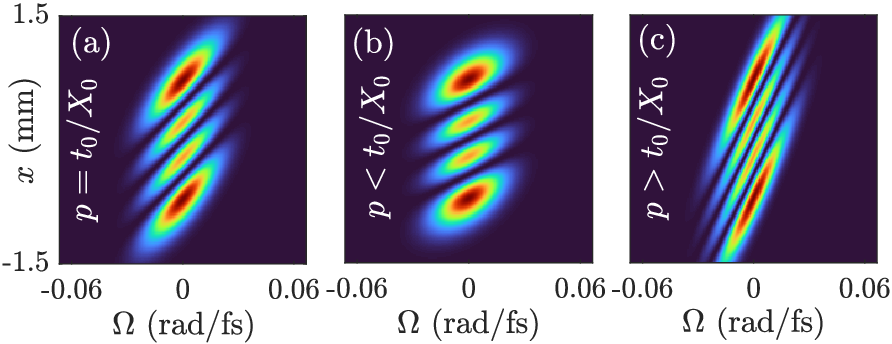}
\caption{Spatially resolved spectral-density of the E-STOV and STOV chains in Fig. \ref{fig:2}. In (b) and (c) we have lowered and enhanced $p$ to $(1/2)t_0/X_0$ and $2 t_0/X_0$ for better visualization of the limits $p\rightarrow 0,\infty$.}
\label{fig:3}
\end{figure}

The spatiospectrum of the E-STOV with $|p|=t_0/X_0$ and maximum spatial chirp is shown in Fig. \ref{fig:3}(a) for comparison with the unmatched cases in Figs. \ref{fig:3} (b) and (c). For decreasing $|p|<t_0/X_0$, as in Fig. \ref{fig:3}(b), the spectral density lobes tend to be horizontal. For increasing $|p|>t_0/X_0$, as in Fig. \ref{fig:3}(c), the lobes tend to be vertical. In both cases the spectral content tends to be the same for all $x$ and hence the spatial chirp tends to disappear. The shapes of spatiospectra underline once again the difference between our E-STOV and standard STOVs. The spatiospectrum of a standard STOV features opposite tilts in its whole structure and in its individual lobes, as the spatiotemporal HG pulse that focuses to a STOV~\cite{porras25}. For E-STOVs, the spatiospectrum is not tilted, but only its lobes.

\section{Transverse orbital angular momentum}\label{sec:OAM}

After Ref.~\onlinecite{porras23-4} and Ref.~\onlinecite{porras24-2}, we can safely evaluate the transverse OAM of the focused TP HG modes. The transverse OAM of the tilted pulse $\psi(x,t)=\psi_x(x)\psi_t(t-px)$ with Eq. (\ref{eq:tilted}) about a transverse $y$ axis crossing the optical axis, and its intrinsic part and extrinsic parts are all null \cite{porras24-1}. The lens splits this null transverse OAM into opposite intrinsic and extrinsic transverse OAM \cite{porras24-1}, and then still null total OAM. We can then restrict ourselves from now on to the intrinsic contribution of the focused TP HG mode. The intrinsic transverse OAM \cite{porras23-4}
\begin{equation}\label{TOAM}
J_y^{(i)} = -\frac{\varepsilon_0c}{2k_0}\int_{-\infty}^{\infty}\int_{-\infty}^{\infty} {\rm Im}\left\{\psi^\star\frac{\partial\psi}{\partial x}\right\} t' dxdt' , 
\end{equation}
($\varepsilon_0$ is the permittivity of vacuum) per unit energy $W=(\varepsilon_0c/2)\int \int|\psi|^2 dxdt'$ of focused generic tilted pulses was evaluated \cite{porras24-1} to be $J_y^{(i)}/W= (p/f)\Delta x^2$, where
\begin{equation}\label{eq:rms}
    \Delta x^2 = \frac{\int_{-\infty}^{\infty} |\psi_x|^2 x^2 dx}{{\int_{-\infty}^{\infty}} |\psi_x|^2 dx}.
\end{equation}
As shown in Appendix \ref{sec:appC}, the last two integrals with the TP HG mode in Eq. (\ref{eq:tilted}) can be performed analytically, resulting in
\begin{equation}\label{TOAM2}
\frac{J_y^{(i)}}{W}=\frac{p}{f} \frac{X_0^2}{4}(2n+1),
\end{equation}
which is proportional to the tilt parameter and the Hermite order $n$.

\begin{figure*}[htb]
\centering
 \includegraphics[width=172mm]{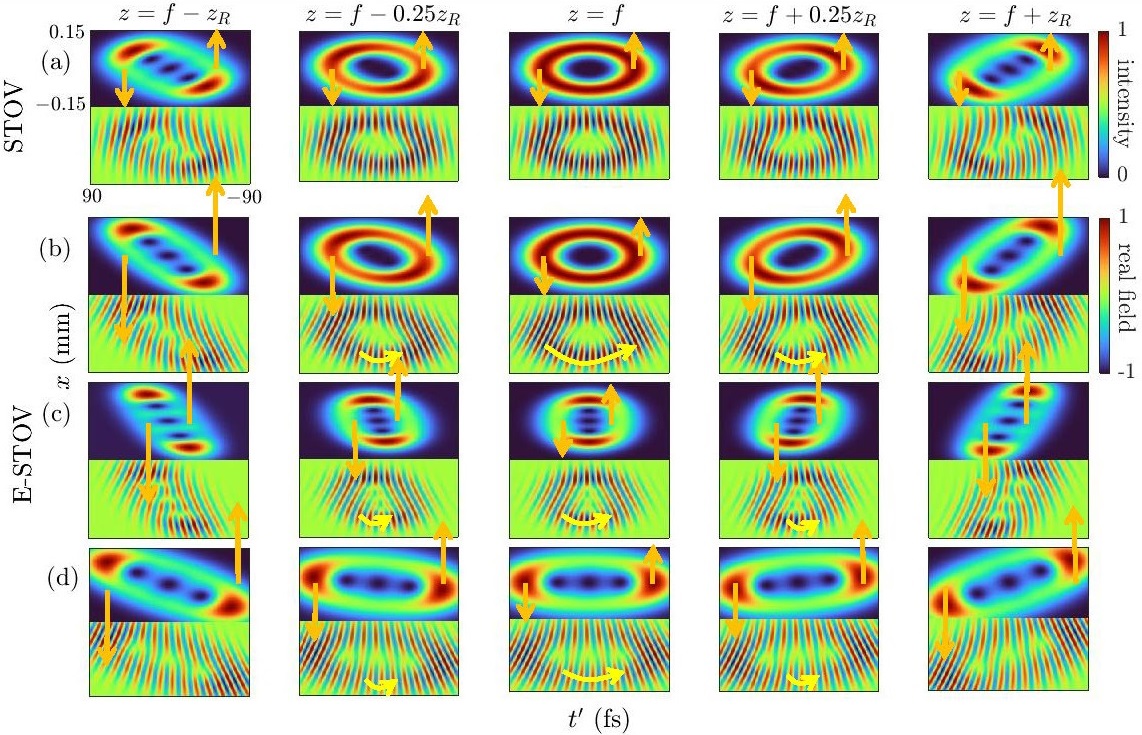}
\caption{(a) Spatiotemporal intensity and real field distributions of a standard STOV at different distances from the focal plane (the carrier frequency is reduced by a factor of 5 for easier visibility). (b,c,d) The same for the E-STOV and STOV chains produced with focused TP HG modes in Fig. \ref{fig:2}(a,b,c). Note that the temporal abscissa is reversed to visualize the rotations as they actually occur in real space $z$-$x$. The transverse OAM points perpendicular to the plane of the figure. $z_R=k_0x_0^2/2$ is the Rayleigh range of the focused field. The orange and yellow arrows indicate opposite momenta and wave front rotation that contribute to the transverse OAM.} 
\label{fig:4}
\end{figure*}

To understand the origin of this intrinsic rotation, we have first plotted in Fig. \ref{fig:4}(a) the focusing on a standard STOV \cite{porras24-1} of the same size $\sqrt{2}x_0$ and duration $\sqrt{2}t_0$ as the E-STOV formed here with $p=\pm t_0/X_0$. The intrinsic transverse OAM of the standard STOV is $J_y^{(i)}/W = \pm  n\gamma/2\omega_0$, where $\gamma=c\sqrt{2}t_0/\sqrt{2}x_0 = ct_0/x_0$ is the ellipticity, and is associated with pulse front rotation~\cite{hyde20} originated by the two opposite, $x$-directed momenta imparted by the lens, drawn as two orange arrows in Fig. \ref{fig:4}(a) \cite{porras23-4}.

For focused TP HG modes, a simple reasoning evidences that part of the transverse OAM imparted by the lens as pulse front rotation transfers to transverse OAM associated with wave front rotation in the vicinity of the focus, as illustrated by the decreasing length of the opposite momenta about the focus and the increasing curved yellow arrows in Fig. \ref{fig:4}(b-d): For E-STOV with $p=\pm t_0/X_0$ in Fig. \ref{fig:4}(b), using that $X_0=2f/k_0x_0$ and introducing $\gamma= c\sqrt{2}t_0/\sqrt{2}x_0=ct_0/x_0$, Eq.  (\ref{TOAM2}) becomes
\begin{equation}\label{TOAM3}
\frac{J_y^{(i)}}{W} = \pm n \frac{\gamma}{2\omega_0} \pm (n+1)\frac{\gamma}{2\omega_0},
\end{equation}
where we have deliberately separated the intrinsic transverse OAM of standard STOVs in the first term.
To track the origin of the second we use the conservation of the transverse OAM to reevaluate $J_y^{(i)}/W$, not immediately after the lens in Eq. (\ref{TOAM2}), but directly from the STOV expression (\ref{eq:elliptical}) conveniently written as $\psi=\psi_{\rm S}e^{\mp i xt'/x_0t_0}$, where $\psi_{\rm S}$ is the standard STOV without the spatial chirp and $e^{\mp i xt'/x_0t_0}$ describes the spatial chirp. Thus, the integrand of the transverse OAM in Eq. (\ref{TOAM}) yields two terms: $\psi^\star \partial_x\psi t'= \psi_{\rm S}\partial_x\psi_{\rm S}t'\mp (t^{\prime 2}/x_0 t_0)|\psi_{\rm S}|^2$. The first term gives the standard STOV contribution $\pm n\gamma/2\omega_0$ in Eq. (\ref{TOAM3}). The second term is then associated with the spatial chirp forming the rotatory wave fronts, which must necessarily give the contribution $\pm (n+1)\gamma/2\omega_0$ at the focal plane. An analogous transfer of transverse OAM from rotating pulse fronts to rotating wave fronts is expected for STOV chains with $p\neq \pm t_0/X_0$ they are also spatially chirped at the focus, as in Figs. \ref{fig:4}(c,d).

As for the amount of transverse OAM, the TP HG modes focus to E-STOVs or STOV chains and defocus faster than the standard STOV, as seen in Figs. \ref{fig:4}(b-d), meaning that the two opposite momenta are bigger. However, for $|p|<t_0/X_0$ [Fig. \ref{fig:4}(c)], the two momenta are close because the tilt is low, thus providing a low transverse OAM. Equating (\ref{TOAM2}) to $\pm n\gamma/2\omega_0$, one finds the $x$-sorted vortex chain that carries the same transverse OAM as a standard STOV as that produced with $|p|= [n/(2n+1)]t_0/X_0< t_0/X_0$. With $|p|=t_0/X_0$ [Fig. \ref{fig:4}(b)] the transverse OAM of the E-STOV is, according to Eq. (\ref{TOAM3}), more than twice that of standard STOVs, and $|p|>t_0/X_0$, the two opposite momenta are further apart, providing even higher transverse OAM for the temporal vortex chain.

\section{Longitudinal field} \label{sec:axial}

Although we only consider paraxial beams in this work, a longitudinal field is nonetheless present. We consider polarization along $x$ since polarization along $y$ is trivial. The longitudinal field can be calculated using the lowest order correction to the paraxial approximation according to Lax {\it et al.} perturbation theory \cite{Lax} as $\psi_z=(i/k_0)(\partial \psi /\partial x)$. We also consider only the waist $z=f$ where the fields are the strongest and most localized. This approximation essentially includes the first-order correction with the small parameter $\epsilon=x_0/z_R$, where the strength of the longitudinal field compared to the transverse field is proportional to $\epsilon$.

For a standard STOV $\psi_{S}\propto \exp(-x^2/x_0^2)\exp(-t'^2/t_0^2)(t'/t_0\mp i x/x_0)^n$, the longitudinal field is
\begin{equation}\label{eq:axialSTOV}
\psi_{S,z}(x,t')=\frac{i\psi_{S}}{k_0}\left[\frac{\mp in/x_0}{\left(\frac{t'}{t_0} \mp i\frac{x}{x_0}\right)} - \frac{2x}{x_0^2}\right].
\end{equation}
For the E-STOV $\psi\propto \exp(-x^2/2x_0^2)\exp(-t'^2/t_0^2)(t/\sqrt{2}t_0\mp i x/\sqrt{2}x_0)^n\exp(\mp i xt'/x_0t_0)$ produced with TP HG mode with input tilt $|p|=t_0/X_0$, the longitudinal field is
\begin{equation}\label{eq:axialSCSTOV}
\psi_z(x,t')=\frac{i\psi}{k_0}\left[\frac{\mp in/\sqrt{2}x_0}{\left(\frac{t'}{\sqrt{2}t_0} \mp i\frac{x}{\sqrt{2}x_0}\right)} - \frac{x}{x_0^2} \mp i\frac{t'}{x_0t_0}\right],
\end{equation}
which will be compared with that of the normal STOV $\psi_S$ of the same size (replacing $t_0\rightarrow \sqrt{2}t_0, x_0\rightarrow \sqrt{2}x_0$) in Eq. (\ref{eq:axialSTOV}), that is, with
\begin{equation}\label{eq:axialSTOV2}
\psi_{S,z}(x,t')=\frac{i\psi_{S}}{k_0}\left[\frac{\mp in/\sqrt{2}x_0}{\left(\frac{t'}{\sqrt{2}t_0} \mp i\frac{x}{\sqrt{2}x_0}\right)} - \frac{x}{x_0^2}\right],
\end{equation}
and with the same replacement in $\psi_S$.

\begin{figure}[ht]
\centering
 \includegraphics[width=70mm]{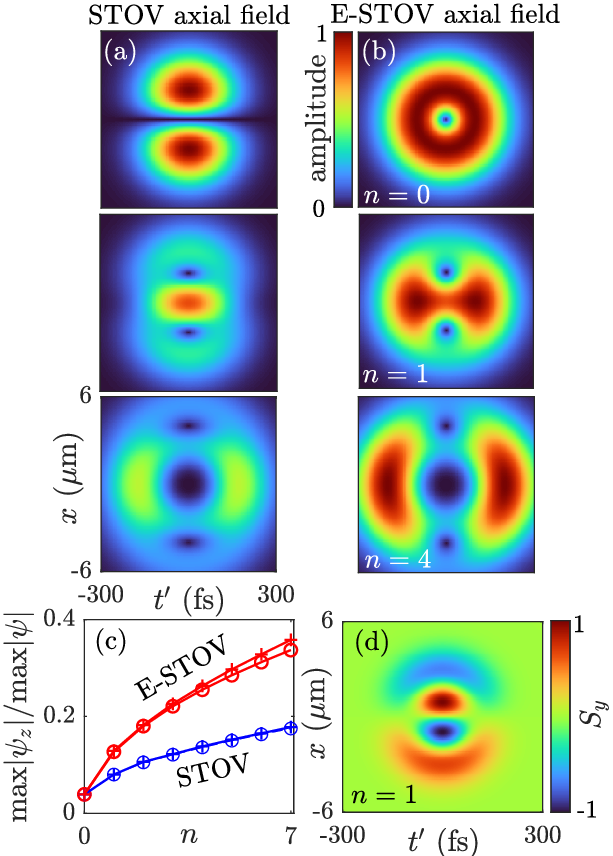}
\caption{Paraxial longitudinal fields evaluated with Eqs. (\ref{eq:axialSCSTOV}) and (\ref{eq:axialSTOV2}) with $t_0=100$ fs, $x_0=2$ $\mu$m for $\lambda_0=800$ nm, corresponding to strongly focusing TP HGs modes, but still with paraxial numerical aperture NA$= \sin[\tan^{-1}(X_0/f)]=\sin[\tan^{-1}(2/k_0x_0)]= 0.126$. Amplitude of the longitudinal field of (a) normal STOV of the same size and (b) E-STOVs for the indicated values of $n$. In each row the amplitude is normalized to the peak amplitude of the E-STOV to visualize the enhanced amplitude for E-STOVs. (c) Ratio longitudinal/transversal peak amplitudes versus $n$ for STOVs (blue) and E-STOVs (red). Circles correspond to the paraxial axial field. Crosses correspond to the nonparaxial axial field.} (d) Transversal SAM density for both STOVs and E-STOVS, as they are identical.
\label{fig:5}
\end{figure}

Figure~\ref{fig:5} compares the longitudinal fields of (a) STOVs and (b) E-STOVs.
The case $n=0$ (the focused field of a tilted Gaussian wave packet) studied in \cite{porras24-1} is included here because, interestingly, the longitudinal field forms a perfect elliptical STOV of charge $\pm 1$ also with the added spatial chirp, i.e., an E-STOV, as is also evident from Eq. (\ref{eq:axialSCSTOV}) with $n=0$. To our knowledge, no longitudinal STOV field has been described so far. This finding calls for an extended, nonparaxial analysis to enhance its amplitude. 

With $n>0$ and for both STOVs and E-STOVs, the longitudinal field features a central vortex of charge $\pm(n-1)$ plus $2$ peripheral, vortices of charge $\pm 1$, all them at $t'=0$. It is also of interest that the longitudinal field grows with $n$, and is significantly larger for E-STOVs than for STOVs, as shown in Fig. \ref{fig:5}(c, circles). Remarkably, the ratio between the longitudinal and transversal peak amplitudes may reach several tens percent for high topological charge and quite tight, but still paraxial focusing, as in the example of Fig. \ref{fig:5}. Numerical simulations have recently been reported of electron acceleration with high-power tightly-focused STOVs \cite{sunF24}. Therefore more work is needed to find the nonparaxial fields necessary for such extreme phenomena~\cite{marceau13-1}. Here, we have just verified that the enhanced longitudinal field in the paraxial fields in Fig. \ref{fig:5} predicted by Lax {\it et al} theory \cite{Lax} are correct by numerical evaluation of these longitudinal fields directly from Gauss's divergence law without any approximation, as detailed in Appendix \ref{sec:appD}. As seen in Fig. \ref{fig:5}(c) the exact prediction (crosses) supports the paraxial approximation (circles). For high topological charge, the axial component is even slightly higher.

Associated with this longitudinal field, there is a local transverse spin angular momentum (SAM) in the spatiotemporal regions where the longitudinal and transversal fields overlap. One can evaluate the SAM density along $y$, i.e. transverse, as $S_y\propto\textrm{Im}\{E_x^*E_z\}$ (ignoring the contribution from the magnetic field), as exemplified in Fig. \ref{fig:5}(d). Interestingly, although the longitudinal fields of STOVs and E-STOVs are clearly different, the transverse SAM density is identical in both cases.

\section{Conclusion}

We have shown that a pulsed, standard Hermite-Gaussian mode acquires a complex spatiotemporal structure when it is tilted and focused. This structure is an elliptical STOV with enhanced properties with regard its spatial chirp, angular momentum and axial component, or an array of transversally or temporally sorted vortices. Closed-form analytical expressions for their formation from the tilted Hermite-Gaussian mode are given. The generation of E-STOVs is simpler compared to the generation of standard STOVs. We only need to tilt and focus standard, spatial HG modes, instead of focusing spatiotemporal HG fields that can only be generated with complex pulse shapers. 

In view of recent advances in high-harmonic-related experiments, the properties of these STOVs make them a promising tool to be used as driving fields in similar experiments. High-harmonic STOVs in the extreme ultraviolet have recently been generated using standard, infrared STOVs as driving pulses \cite{rodrigo24}. In addition, these new STOVs incorporate the wave front rotation at the focal plane that enables the lighthouse effect for the generation of isolated attosecond pulses \cite{vicenti12,quere14,auguste16}. Together, these results suggest that the use of E-STOVs as drivers may result in the generation of isolated, attosecond STOV-like structures.  

We have also examined the longitudinal field imposed by Gauss' divergence law in the paraxial approximation, and described for the an elliptical STOV in the longitudinal field whose  transversal field is a rotating pulse without any phase singularity. STOVs of higher order in the longitudinal field are not so perfect, but their amplitude is larger than that of standard STOVs, and increases with the topological charge. In view of the foreseen applications of STOVs in electron trapping and acceleration \cite{sunF24}, more work is needed to describe and realize these longitudinal STOV fields at amplitude level comparable to that of the transversal field in nonparaxial focusing.   

\begin{acknowledgments}
Fonds De La Recherche Scientifique - FNRS. Ministerio de Ciencia e Innovación (PID2021-122711NB-C21).
\end{acknowledgments}

\section*{Author declarations}

\subsection*{Data availability}
The data that supports the findings of this study are available within the article.

\subsection*{Conflict of interest}
The authors have no conflicts to disclose.

\subsection*{Author Contributions}
M.A.P. and S.W.J. contributed equally to this work.

\appendix

\section{Fresnel diffraction integral for paraxial focusing of quasimonochromatic pulses and derivation of Eq. (\ref{eq:main})}
\label{sec:appA}

Starting with a scalar field satisfying the wave equation $\Delta E - (1/c^2)\partial^2E/\partial t^2=0$, we change variables to $t'=t-z/c$ and $z'=z$ to obtain
\begin{equation}
\Delta_\perp E + \frac{\partial^2 E}{\partial z^{\prime 2}} - \frac{2}{c}\frac{\partial^2 E}{\partial z'\partial t'}=0,
\end{equation}
where $\Delta_\perp$ is the transverse Laplacian.
With $E= \psi (x,y,t',z')e^{-i\omega_0 t'}$, the equation for the envelope is
\begin{equation}
\Delta_\perp \psi + \frac{\partial}{\partial z'}\left[\frac{\partial \psi}{\partial z'} - \frac{2}{c}\left(\frac{\partial\psi}{\partial t'}-i\omega_0\psi\right)\right]=0.
\end{equation}
For paraxial and quasimonochromatic pulses the axial variation of the envelope is much slower than the axial oscillations, $|\partial \psi/\partial z'|\ll k_0|\psi|$, and its temporal variation is much slower than the temporal oscillations, $|\partial \psi/\partial t'|\ll \omega_0 |\psi|$, which simplifies the above equation to
\begin{equation}
\Delta_\perp \psi + 2ik_0 \frac{\partial\psi}{\partial z'} =0,   
\end{equation}
where the local time does not appear explicitly, and is therefore formally identical to the paraxial wave equation for monochromatic light beams. Its solution with initial $\psi(x,y,t')$ is given by Fresnel diffraction integral. For factorized $\psi(x,t')\phi(y,t')$, each factor satisfies the 1D Fresnel diffraction integral
\begin{equation}
\psi(x,t',z') = \sqrt{\frac{k_0}{2i\pi z' }}\int_{-\infty}^{\infty} dx' \psi(x',t')e^{\frac{ik_0}{2z}(x-x')^2}\,. 
\end{equation}
When an ideal lens of focal length $f$ is illuminated by the field $\psi(x,t')$ a wave front and pulse front curvatures are introduced, expressed by $\psi(x,t'-x^2/2cf)e^{-ik_0 x^2/2f}$. If however, the maximum temporal delay $X_0^2/2cf$ of a pulse of transversal size $X_0$ due to pulse front curvature is much smaller than the pulse duration $t_0$, the effect of pulse front curvature is small or negligible, leading to
\begin{equation} \label{eq:Fresnel2}
\psi(x,t',z') = \sqrt{\frac{k_0}{2i\pi z' }}\int_{-\infty}^{\infty} dx' \psi(x',t')e^{-\frac{ik_0x^{\prime 2}}{2f}}e^{\frac{ik_0}{2z}(x-x')^2}\,, 
\end{equation}
for focusing under the condition $X_0^2/2cf \ll t_0$.

\begin{figure}[htb]
\centering
 \includegraphics[width=86mm]{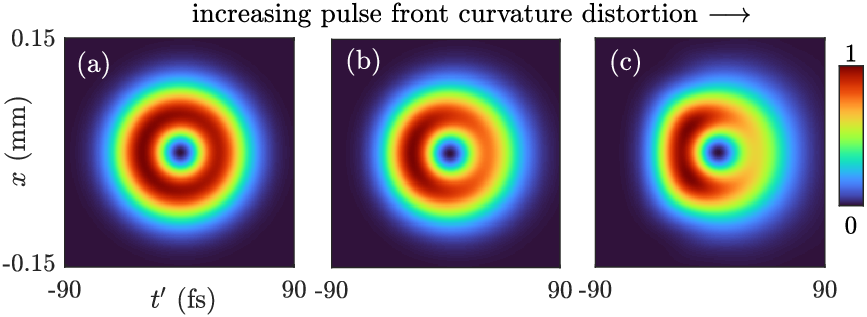}
\caption{Distortion of the focal E-STOV (intensity) with $x_0=0.051$ mm, $t_0=30$ fs, $n=1$ at $\lambda_0=800$ nm created by focusing TP HG modes as $f$ approaches the upper bound $720$ mm in (\ref{eq:focal2}) for small pulse front curvature effects. (a) $f=50$ mm, $X_0=0.25$ mm, (b) $f=100$ mm, $X_0=0.5$ mm, (c) $f=200$ mm, $X_0=1$ mm. They are calculated numerically using $\psi(x',t'-x^2/2cf)e^{-ik_0 x^2/2f}$ in Fresnel integral (\ref{eq:Fresnel2}), i.e., without neglecting pulse front curvature. All amplitudes are normalized to their peak values.}
\label{fig:6}
\end{figure}

In practice, one may wish to produce the fields described in this paper from Eq. (\ref{eq:Fresnel2}) with a given focal size $x_0= 2f/k_0X_0$ and duration $t_0$. Then, the condition $X_0^2/2cf \ll t_0$ of negligible pulse front curvature effects leads to the condition
\begin{equation} \label{eq:focal2}
    f\ll ct_0k_0^2x_0^2/2 
\end{equation}
for the focal length, and accordingly $X_0=2f/k_0x_0$ for the input size. Figure \ref{fig:6} is evaluated numerically without neglecting the effects of pulse front curvature. A nearly perfect E-STOV is formed when $f$ amply satisfies condition (\ref{eq:focal2}), and becomes increasingly distorted when $f$ approaches the upper bound in (\ref{eq:focal2}).

\textbf{Derivation of Eq.~(\ref{eq:main}):} Equation (\ref{eq:Fresnel}) with $\psi_x(x')\psi_t(t'-px')$ given by Eqs. (\ref{eq:tilted}) can be written as 
\begin{equation}\label{eq:inter}
\psi(x,t',z)= A\int_{-\infty}^{\infty} dx' e^{-x^{\prime 2}/X_{0, \rm eff}^2}e^{2ax'}H_n\left(\frac{\sqrt{2}x'}{X_0}\right),    
\end{equation}
where $X_{0,\rm eff}$ is defined in Eq. (\ref{eq:xeff}), 
\begin{equation}
    a= \left(\frac{pt'}{t_0^2}-\frac{ik_0x}{2z}\right),
\end{equation}
and 
\begin{equation}
    A= \sqrt{\frac{k_0}{2i\pi z}}\frac{1}{2^n} e^{-\frac{t^{\prime 2}}{t_0^2}}e^{\frac{ik_0 x^2}{2z}}.
\end{equation}
Completing the square, Eq.~(\ref{eq:inter})
transforms into
\begin{equation}
    \psi(x,t',z) = Ae^{a^2X_{0,\rm eff}^2}\int_{-\infty}^{\infty} dx' e^{\left(\frac{x'}{X_{0,\rm eff}}-aX_{0,\rm eff}\right)^2}H_n\left(\frac{\sqrt{2}x'}{X_0}\right),
\end{equation}
which can be immediately identified with integral 7.374.8 of Ref.~\onlinecite{gradshteyn07}, leading to our main result in Eq. (\ref{eq:main}). 

\section{Frequency-domain description}
\label{sec:appB}

The time-domain derivation in the main text is extremely intuitive and captures the formation of E-STOVs and STOV chains from focusing TP HG modes. As seen in Eq.~(\ref{eq:spatiospectrum}) for the spatiospectrum at the focal plane, there is an alternate description in the frequency-domain starting with standard Hermite-Gauss modes of different frequencies with a spatial chirp along $x$, represented by the replacement $x \rightarrow x-b\Omega$, where $\Omega=\omega-\omega_0$. Next we can use the well-known propagation relations for monochromatic HG modes to reconstruct the spatiospectrum at any distance from the focus. With this strategy, Eq.~(\ref{eq:spatiospectrum}) becomes 
\begin{align}
\begin{split}
\hat{\psi}(x,\Omega, z)=&e^{\frac{-\Omega^2}{\Omega_0^2}}\sqrt{\frac{-iz_R}{q}}\left(\frac{q^*}{q}\right)^{\frac{n}{4}}e^{\frac{i\omega(x-b\Omega)^2}{2cq}}\\
&\times\frac{1}{2^n}H_n\!\left[\frac{\sqrt{2}(x\!-\!b\Omega)}{w}\right]
\label{eq:HG_SC_freq},
\end{split}
\end{align}
assuming for simplicity that $z=0$ is now the focal plane, $q=z-iz_R$, $w=x_0\sqrt{1+(z/z_R)^2}$, and $z_R=k_0 x_0^2/2$. The result in time-domain can be re-evaluated ignoring a term $\propto\Omega^3$ in an exponential in Eq.~(\ref{eq:HG_SC_freq}), and inverse Fourier transform using integral 7.374.8 in Ref.~\onlinecite{gradshteyn07}, yielding
\begin{align}
\begin{split}
&\hat{\psi}(x,\Omega, z)=\frac{1}{\alpha\Omega_0}
\sqrt{\frac{-iz_R}{q}}\left(\frac{q^*}{q}\right)^{\frac{n} {4}}e^{\frac{i\omega_0 x^2}{2cq}} e^{-\frac{t^{\prime\prime 2}}{4\alpha^2}} \\
&\times \left(1-\frac{2b^2}{\alpha^2w^2}\right)^{\frac{n}{2}} \frac{1}{2^n}H_n\left(\frac{\sqrt{2}}{w}
\frac{x+
\frac{ibt^{\prime\prime}}{2\alpha^2}}{\sqrt{1-\frac{2b^2}{\alpha^2w^2}}}\right)
\label{eq:HG_SC_time},
\end{split}
\end{align}
with the parameters
\begin{align}
\alpha&=\sqrt{\frac{1}{\Omega^2_0}+\frac{i(2xb-\omega_0b^2)}{2cq}}\approx\sqrt{\frac{1}{\Omega^2_0}-\frac{i\omega_0b^2}{2cq}},\\
t^{\prime\prime}&=t'-\frac{x^2}{2cq}+\frac{\omega_0 b x}{cq}\approx t'-\frac{x^2z}{2c(z^2+z_R^2)}+\frac{\omega_0 b x}{cq}.
\end{align}
This is the same procedure as in our recent work with vortex beams~\cite{porras23-3} and vector beams~\cite{jolly24-1} having spatial chirp. With lengthy algebra this can be confirmed to be  equivalent to Eq.~(\ref{eq:main}), aside a global constant factor, when $z$ is replaced with $z+f$ in that equation, and identifying $p=(k_0/f)b$ and $\Omega_0=2/t_0$. This method does not include the curvature factor $e^{ik_0 x^2/2f}$ (which is negligible with tight focusing), but it does model the delay due to pulse-front curvature seen in the term $\propto x^2$ in $t^{\prime\prime}$, which is important when the conditions such as (\ref{eq:focal2}) are not satisfied. Therefore, this technique is particularly suited to include time delay due to pulse-front curvature. At the waist ($z=f$ in Eq.~(\ref{eq:main}), i.e. Eq.~(\ref{eq:focal}), and at $z=0$ in Eq.~(\ref{eq:HG_SC_time}), i.e. Eq.~(\ref{eq:spatiospectrum}), these equations are the same aside from a constant, as already confirmed in the main text.

This frequency domain view once again underscores the simplicity of our scenario, where the E-STOV is a spatially-chirped Hermite-Gaussian mode, but the standard STOV is a more complex field requiring pulse shaping before focusing.

\section{Derivation of Eq. (\ref{TOAM2}) for the transverse OAM}
\label{sec:appC}

The intrinsic transverse OAM per unit energy of a focused tilted pulse is given by Ref.~\onlinecite{porras24-1} $J_y^{(i)}/W =(p/f) \Delta x^2$, with
\begin{equation}
    \Delta x^2 = \frac{\int_{-\infty}^{\infty} |\psi_x|^2 x^2 dx}{\int_{-\infty}^{\infty} |\psi_x|^2 dx} =\frac{I_1}{I_2}.
\end{equation}
To evaluate the integrals $I_1$ and $I_2$ with $\psi_x=H_n(\sqrt{2}x/X_0)e^{-x^2/X_0^2}$, we introduce $\alpha=\sqrt{2}x/X_0$ and use that $\alpha^2 = H_2(\alpha)/4 + H_0(\alpha)/2$. The integrals $I_1$ and $I_2$ become
\begin{eqnarray}
I_1 &=&\frac{X_0^3}{2\sqrt{2}}\int_{-\infty}^{\infty} H_n^2(\alpha)\left(\frac{1}{4}H_2(\alpha)+ \frac{1}{2}H_0(\alpha)\right)e^{-\alpha^2}d\alpha, \nonumber \\
I_2 &=&\frac{X_0}{\sqrt{2}}\int_{-\infty}^{\infty} H_n^2(\alpha)H_0(\alpha)e^{-\alpha^2}d\alpha , \nonumber 
\end{eqnarray}
which can be performed using integrals 7.375.2 in Ref.~\onlinecite{gradshteyn07}, yielding
\begin{eqnarray}
I_1 &=&\frac{X_0^3}{2\sqrt{2}}\sqrt{\pi}2^n n!(n+1/2) ,\nonumber \\
I_2 &=& \frac{X_0}{\sqrt{2}} \sqrt{\pi}2^n n!. \nonumber 
\end{eqnarray}
Their quotient is $\Delta x^2 = (X_0^2/4)(2n+1)$, and the transverse OAM per unit energy
\begin{equation}
\frac{J_y^{(i)}}{W}=\frac{p}{f} \frac{X_0^2}{4}(2n+1).
\end{equation}

\section{Verification of the enhanced longitudinal component with Gauss's divergence law}
\label{sec:appD}

The longitudinal component is fully determined by the transversal components by Gauss's divergence law $\nabla\cdot {\bf E}=0$ in free space, or for $x$ polarization, $\partial E_x/\partial x + \partial E_z/\partial z=0$. For the spatial Fourier transform
divergence law reads $k_x \hat E_x(k_x,t')+ k_z \hat E_z(k_x,t') =0$, where $k_z=\sqrt{k_0^2-k_x^2}$. Thus
\begin{equation}
    E_z(x,t') = -FT^{-1}\left(\frac{k_x\hat E_x(k_x,t')}{\sqrt{k_0^2-k_x^2}}\right),
\end{equation}
where $FT^{-1}$ means inverse Fourier transform. Since $E_i(x,t') =\psi_i(x,t')e^{-i\omega_0 t'}$, $i=x,z$, the envelopes also verify
\begin{equation}\label{eq:gauss}
    \psi_z(x,t') = -FT^{-1}\left(\frac{k_x\hat \psi_x(k_x,t')}{\sqrt{k_0^2-k_x^2}}\right).
\end{equation}
In Sec. \ref{sec:axial} we have applied Eq. (\ref{eq:gauss}) for the STOV, $\psi_x=\psi_S$ and the E-STOV $\psi_x=\psi$. When $k_x$ in the denominator of (\ref{eq:gauss}) is neglected, we recover the paraxial approximation $\psi_z =(i/k_0)\partial\psi_x/\partial x$.

\bibliography{HG-SC_biblo}

\end{document}